\documentclass[aps,pre,twocolumn,groupedaddress,showpacs]{revtex4}
\usepackage{graphicx}
\usepackage{dcolumn}
\usepackage{bm}
\usepackage{epsfig}
\usepackage{color}
\usepackage{float}
\input epsf
\epsfclipon

\begin{document}
\title{Growing spin model in deterministic and stochastic trees} 
\author{Julian Sienkiewicz}
\affiliation{Faculty of Physics, Center of Excellence for Complex Systems Research, Warsaw University of Technology, Koszykowa 75, PL-00-662 Warsaw, Poland}
\date{\today}
\begin{abstract}
We solve the growing asymmetric Ising model [Phys. Rev. E {\bf 89}, 012105 (2014)] in the topologies of deterministic and stochastic (random) scale-free trees predicting its non-monotonous behavior for external fields smaller than the coupling constant $J$. In both cases we indicate that the crossover temperature corresponding to maximal magnetization decays approximately as $(\ln \ln N)^{-1}$, where $N$ is the number of nodes in the tree.
\end{abstract} 
\maketitle

\section{INTRODUCTION}

Although one-dimensional systems can be often used to model social dynamics \cite{sznajd, rumor, isolation}, it is usually observed that the structure of the majority of online social systems such portals or fora follows a different type of topology --- a scale-free one that is reflected in their degree distribution \cite{ba,huberman}. Such non-trivial topologies have motivated several researchers to explore the behavior of the one of the most fundamental approach of the statistical physics --- Ising model \cite{ising}, which has been tested on Cayley trees \cite{cal}, BA networks \cite{agata} or growing trees \cite{has}, to mention a few. 

However, there is no direct evidence that social processes that take place in hierarchical trees and scale-free networks can be described by this kind of dynamics. On the other hand the results of our previous analyses \cite{plos,ania} indicate that one of the most dominant phenomena seen in online portals is a strong dependence of the expressed emotion on the emotion of the last comment (i.e., the newest one). In order to describe this process in setting of chronologically added comments (that form a chain) we have previously explicitly modified Ising model Hamiltonian by taking into account only node's left neighbor as well as equip our model with a growing component (a new node is quenched after a single update) \cite{eising}. Here, we extend this concept to tree topologies (deterministic and stochastic) in order to explore the influence exerted by intrinsic features of those systems onto the behavior of the model.   

The paper is organized as follows: in Sec. \ref{sec:model} we describe in short of the basic concepts of the model introduced in \cite{eising}. Sections \ref{sec:tree} and \ref{sec:sf} gather the results obtained applying dynamics to deterministic and stochastic scale-free trees, respectively. Finally Sec. \ref{sec:con} concludes the paper discussing differences between the considered topologies and growing chain.

\section{MODEL DESCRIPTION}\label{sec:model}

The basic version of the model uses the idea of a growing chain: the first node of the chain has a random spin $s_0=\pm 1$ (it can be interpreted as an emotional valence \cite{valence} of a post in online discussion), drawn with probability $\Pr(s_0= \pm 1)=1/2$. After that, another node of the chain is added to the right side of the last one and it is initially equipped with a spin once again drawn with equal probabilities $\Pr(s_1= \pm 1) = 1/2$. In the following step, the node becomes a subject to the updating procedure that is based on the Ising-like model approach. For each appearing node $n$ we define a function $\mathcal{E}_n = - J s_{n-1} s_n - h s_n$, where the constant $J > 0$ corresponds to exchange integral in the Ising model and $h$ is the external field. The function $\mathcal{E}_n$ can be treated as a type of an {\it emotional discomfort} function felt by a user posting a message $s_n$. After the spin is drawn, we check how flipping its sign to the opposite one (i.e., from $s_n=+1$ to $s_n=-1$ or likewise) affects the change of function $\mathcal{E}$ as $\Delta \mathcal{E} = \mathcal{E}'_n - \mathcal{E}_n = - (Js_{n-1} + h)(s'_n-s_n)$, where term $\mathcal{E}'_n$ corresponds to $s_n'$ calculated when $s_n \rightarrow s'_n =-s_n$. Then we follow the Metropolis algorithm \cite{metropolis} i.e., if the $\Delta \mathcal{E}< 0$ we accept the change, otherwise we test if the expression $\exp[-\Delta \mathcal{E}(k_B T)^{-1}]$ is smaller or larger than a random value $\xi \in [0; 1]$ (here $k_B$ is Boltzmann constant and $T$ is temperature). If the latter occurs we accept the change, otherwise the spin is kept as originally chosen. The procedure of adding new nodes and setting their spins is repeated until the size $N$ of the chain is reached. 

It can be shown \cite{eising} that system dynamics follows a two-state Markov chain approach defined by the transition matrix $\mathbf{P}$
\begin{equation}\label{eq:p}
\mathbf{P}=\left[ \begin{array}{cc}
p & 1-p \\
1-q & q
\end{array} \right]
\end{equation}
with conditional probabilities (which come from the above described dynamics) given by 
\begin{equation}\label{eq:hsmall}
\left\{
\begin{array}{l}
p = \Pr \left(+|+\right) = 1-\frac{1}{2} \mathrm{e}^{-\widetilde{\beta}(h+J)}\\
q = \Pr \left(-|-\right) = \frac{1}{2} \pm \frac{1}{2} \mp \frac{1}{2} \mathrm{e}^{\pm \widetilde{\beta}(h-J)}\\
\end{array}
\right.,
\end{equation}
where upper signs correspond to case $|h| < J$, lower signs to $|h| \ge J$ and $\tilde\beta=2/(k_B T)$.

As a result the average spin (or valence) in the $n$th node of the chain is given as
\begin{equation}\label{eq:sn}
\langle s_n \rangle = \frac{p-q}{2-p-q}\left[ 1 - (p+q-1)^n \right],
\end{equation}
while the average spin in the whole chain can be obtained as a mean value over $\langle s_n \rangle$, i.e.,
\begin{equation}
\langle s \rangle = \frac{p-q}{1-Q}\left[1 + \frac{1}{N} - \frac{1 - Q^{N+1}}{N(1- Q)}\right],
\end{equation}
where $Q = p + q - 1$.

The motivation is to compare the results obtained for a chain topology with the ones that are derived for deterministic trees and random scale-free trees.

\section{DETERMINISTIC TREES}\label{sec:tree}
\begin{figure}
\centering
\includegraphics[width=.8\columnwidth]{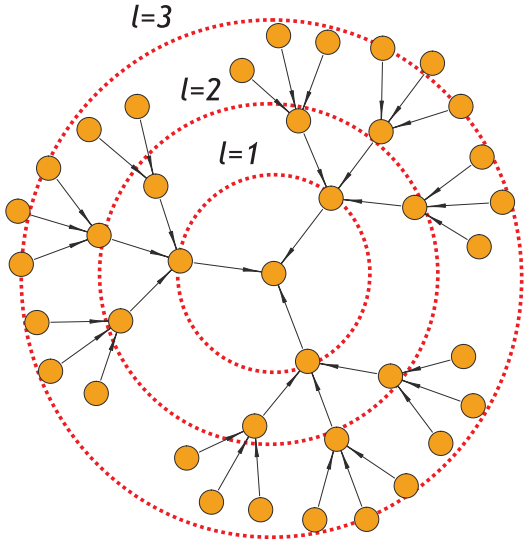}
\caption{(color online) A schematic plot of a deterministic tree with $z=3$ and $L=3$.}
\label{fig:etree}
\end{figure}

The topology of a deterministic tree is described by two parameters: the number of children $z$ each node gives birth to and the depth of the tree $L$. The total number of vertices (expect the root one) is equal to
\begin{equation}\label{eq:nz}
N = \sum_{l=1}^{l=L} z^l = z \frac{z^L-1}{z-1}.
\end{equation}
Figure \ref{fig:etree} illustrates an example of a tree with $z=3$ and $L=3$.

\begin{figure}
\centering
\includegraphics[width=0.5\columnwidth]{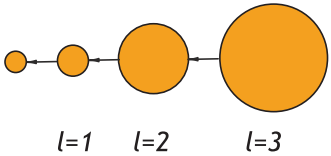}
\caption{(color online) Graphical representation of weighting of spins at a given sept of the tree.}
\label{fig:wchain}
\end{figure}

\begin{figure}
\centering
\includegraphics[width=\columnwidth]{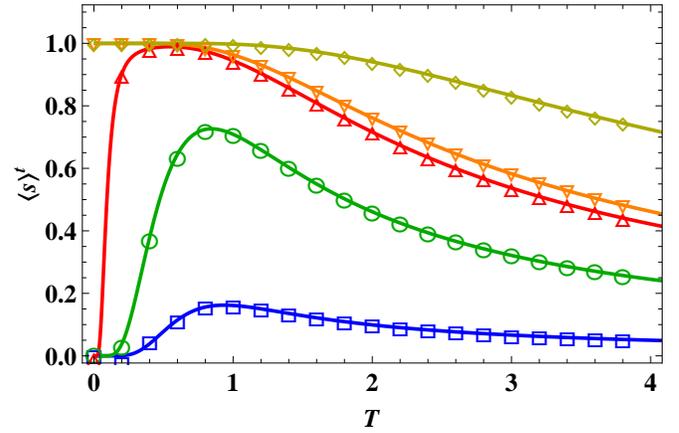}
\caption{(color online) Average spin $\langle s \rangle^t$ in a tree ($z=3$, $L=12$) as function of temperature for different values of the external magnetic field: $h=0.1$ (squares), $h=0.5$ (circles), $h=0.9$ (upward triangles), $h=1$ (downward triangles) and $h=2$ (diamonds). Solid lines come form Eqs. (\ref{eq:sts}) and (\ref{eq:stl}). All data points have been averaged over $M=10^4$ realizations.}
\label{fig:sT_tree}
\end{figure}

The first and key observation one needs to make is that in the case of described model (Sec. \ref{sec:model}) a directed tree can be regarded as equivalent to a chain of length $L$. However, on each level $l$ of the tree there is a different number of nodes that have to be taken into account. As a result one obtains a chain whose spin values should be {\it weighted} by the number of nodes present at a given level (depth) $l$ (see Fig. \ref{fig:wchain}). Then, in order to obtain the formula for the average spin in the tree $\langle s \rangle^t$ one needs to perform the following summation
\begin{equation}
\langle s \rangle^t = \sum_{l=1}^{l=L} \langle s_l \rangle z^l,
\end{equation}
where $\langle s_l \rangle$ is given by Eq. \ref{eq:sn}.

\begin{figure*}
\centering
\begin{tabular}{cc}
\includegraphics[width=\columnwidth]{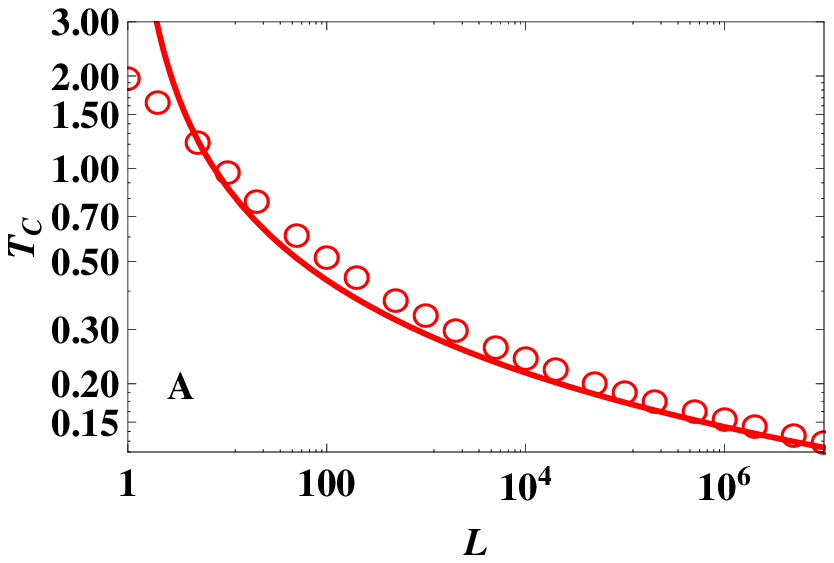} &
\includegraphics[width=\columnwidth]{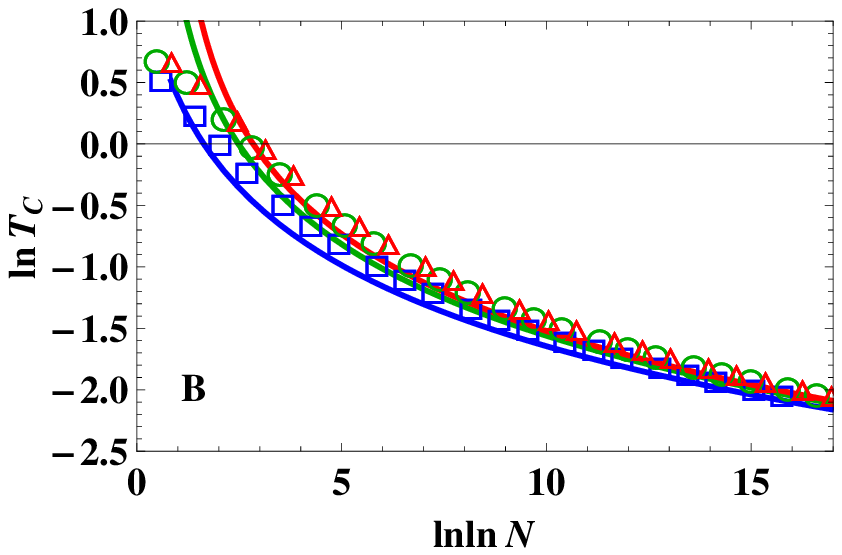}\\
\end{tabular}
\caption{(color online) (a) Crossover temperature $T_c$ versus the depth of the tree $L$. Symbols are numerical solutions of Eq. (\ref{eq:sts}) while the solid line come from $T_c = 2(\ln L)^{-1}$. In all cases $z=10$. (b) The logarithm of the crossover temperature $T_c$ versus the double logarithm of the number of nodes in a tree $N$. Symbols (squares --- $z=2$, circles --- $z=5$, triangles --- $z=10$) are numerical solutions of Eq. (\ref{eq:sts}) while solid lines come from Eq. (\ref{eq:ttc}).}
\label{fig:tcLN}
\end{figure*}

After some algebraic calculations one arrives at the following expression
\begin{equation}
\langle s \rangle^t = \frac{p-q}{1-Q}\left[1 + \frac{1}{N} - \frac{1 - (zQ)^{L+1}}{N(1- zQ)}\right],
\end{equation}
which can be expressed explicitly for $|h| < J$ as
\begin{eqnarray}\label{eq:sts}
\langle s \rangle^t_s & = & \tanh  \widetilde{\beta} h \nonumber \\
& & \times \left\{ 1 + \frac{1}{N} - \frac{1 - \left[\left( 1- \mathrm{e}^{- \widetilde{\beta} J} \cosh  \widetilde{\beta} h \right) z \right]^{L+1}}{N \left[1-\left(1 - \mathrm{e}^{- \widetilde{\beta} J} \cosh  \widetilde{\beta} h \right) z \right]} \right\}
\end{eqnarray}
and for $|h| \ge J$ as
\begin{eqnarray}\label{eq:stl}
\langle s \rangle^t_l & = & \mathrm{sgn}(h) \frac{\cosh \widetilde{\beta} J - \mathrm{e}^{ \widetilde{\beta} |h|}}{\sinh  \widetilde{\beta} J - \mathrm{e}^{ \widetilde{\beta} |h|}} \nonumber \\
& & \times \left[ 1 + \frac{1}{N} - \frac{1 - \left(z \mathrm{e}^{- \widetilde{\beta} |h|} \sinh  \widetilde{\beta} J \right)^{L+1}}{N \left(1 - z \mathrm{e}^{- \widetilde{\beta} |h|} \sinh  \widetilde{\beta} J  \right)} \right]
\end{eqnarray}

As can be seen in Fig \ref{fig:sT_tree} (for simplicity this plot and further ones are for $J = k_B = 1$) the above functions follow a shape that is similar to the one observed for the chain --- we have a maximum in $\langle s(T) \rangle$ for $|h| < J$ and an absence of such behavior for $|h| \ge J$. Having that in mind it is interesting to examine the dependence of the crossover temperature $T_c$ (i.e., the temperature for which $\langle s \rangle^t_s$ takes the maximum) as a function of tree parameters. It can be shown that $T_c \approx 2(\ln L)^{-1}$ (see Appendix \ref{app:a} and Fig. \ref{fig:tcLN}a), which using Eq. (\ref{eq:nz}) and assuming $z^L \gg 1$ and $z \approx z - 1$ gives
\begin{equation}\label{eq:ttc}
T_c \approx \frac{2}{\ln\ln N - \ln \ln z}
\end{equation} 
A comparison of the crossover temperature obtained by numerically solving Eq. (\ref{eq:sts}) with the predictions of Eq. \ref{eq:ttc} is shown in Fig \ref{fig:tcLN}b. For sufficiently large values of $N$ all data points and analytical curves collapse, indicating lack of dependence on parameter $z$.

\section{RANDOM SCALE-FREE TREES}\label{sec:sf}
Although the tree topology (and especially the underlying branching process) is quite common in the real-world, it is rather unreasonable to believe that social media systems follow this construction. According to previous studies one expects that the growth of such systems could be governed by the preferential process \cite{barmp}. In the case of the tree topology this assumption means that we are dealing with the scale-free tree structure, i.e., an evolving system, where in each time step a new node is most likely attached to the one characterized with the highest degree. By following such a procedure one ends up with a tree whose degree distribution is power-law.

\begin{figure*}
\centering
\begin{tabular}{cc}
\includegraphics[width=\columnwidth]{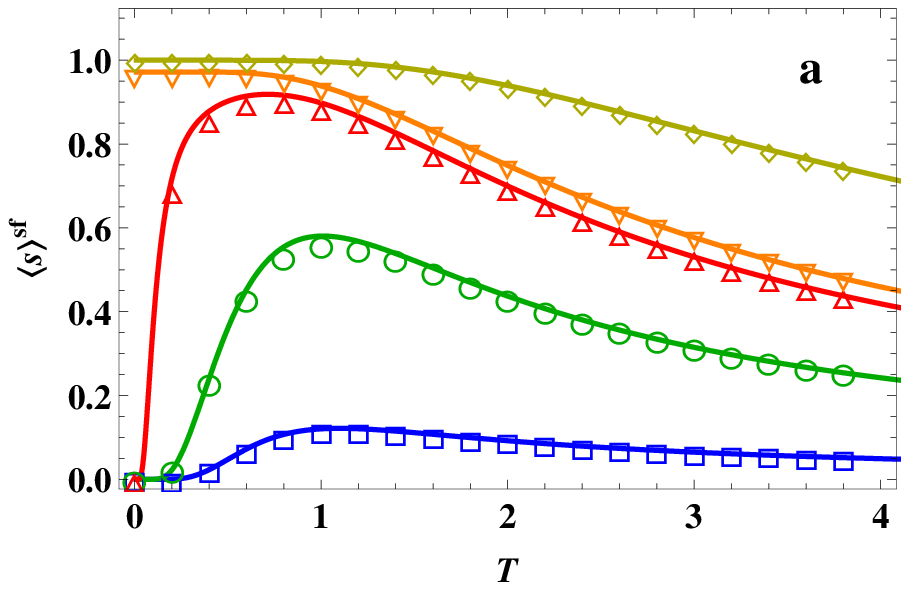} &
\includegraphics[width=\columnwidth]{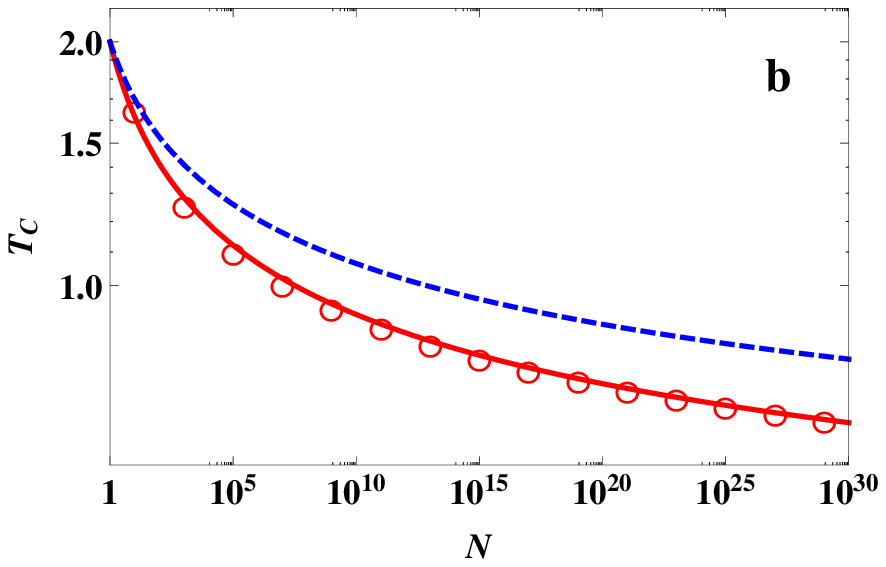}\\
\end{tabular}
\caption{(color online) (a) Average spin $\langle s \rangle^{sf}$ in a random scale-free tree ($N = 10^5$) as function of temperature for different values of the external magnetic field: $h=0.1$ (squares), $h=0.5$ (circles), $h=0.9$ (upward triangles), $h=1$ (downward triangles) and $h=2$ (diamonds). Solid lines come form Eqs. (\ref{eq:ssfs}) and (\ref{eq:ssfl}). All data points have been averaged over $M=10^5$ realizations. (b) Crossover temperature $T_c$ versus the number of nodes $N$. Symbols are numerical solutions of Eq. (\ref{eq:ssfs}), the solid line comes from Eq. (\ref{eq:sftc}) while the dashed line comes form Eq. (\ref{eq:b5}) (see Appendix \ref{app:b} for details).}
\label{fig:ssf}
\end{figure*}

The scheme for obtaining the average spin value in random scale-free trees is similar to the one presented in the previous Section. However, in this case, we are dealing with a stochastic process also in the case of tree formation (i.e., not only dynamics but also the topology). We use the results of Bollob\'{a}s and Riordan \cite{bollobas} and Szab\'{o} {\it et al.} \cite{szabo} that give the mean-field number of vertices $n(l)$ at distance $l$ from the root node
\begin{equation}\label{eq:nl}
n(l) = A \frac{\left(\ln N / 2 \right)^{l-1}}{(l-1)!},
\end{equation}
where $A$ is the number of children of the root node. We assume that the root node possesses the highest degree in the network, thus $A = \sqrt{N}$ \cite{barmp}.
As a consequence, the formula for the average spin in scale-free random tree is given as
\begin{equation}
\langle s \rangle_{sf} = \sum_{l=1}^{l=L} \langle s_l \rangle n(l),
\end{equation}
which results in
\begin{equation}\label{eq:ssf}
\langle s \rangle_{sf} = \frac{p-q}{1-Q} \frac{\Gamma\left( L, \frac{\ln N}{2} \right) - N^{\frac{Q-1}{2}}Q\Gamma\left( L,Q \frac{\ln N}{2} \right)}{\Gamma(L)},
\end{equation}
where $\Gamma(x)$ is gamma function and $\Gamma(a,x)$ is incomplete gamma function. In order to obtain an equation that has only one free parameter connected to topology (i.e., the number of vertices) one needs to calculate tree depth $L$. To do this we use Eq. (\ref{eq:nl}), setting $n(L)=1$. Then, taking the logarithm of both sides and implementing Stirling formula we have
\begin{equation}
(L-1)\ln(L-1) - L+1 \approx \ln \sqrt{\frac{N}{2\pi(L-1)}} + (L-1)\ln \left(\frac{\ln N}{2} \right).
\end{equation}
After omitting the first term on the r.h.s. we get
\begin{equation}
L = 1 + \frac{\mathrm{e}}{2}\ln N
\end{equation}
Thus, Eq. (\ref{eq:ssf}) can be expressed explicitly for $|h| < J$ as
\begin{eqnarray}\label{eq:ssfs}
\langle s \rangle^{sf}_s & = & \tanh  \widetilde{\beta} h \left[ \frac{\Gamma\left( L, \frac{\ln N}{2} \right)}{\Gamma(L)} - \right. \nonumber \\
& & \left. \frac{(1-\mathrm{e}^{-\widetilde{\beta} J} \cosh  \widetilde{\beta} h)\Gamma\left(L,(1-\mathrm{e}^{-\widetilde{\beta} J} \cosh  \widetilde{\beta} h) \frac{\ln N}{2} \right)}{N^{\frac{1}{2}\mathrm{e}^{-\widetilde{\beta} J} \cosh  \widetilde{\beta} h}\Gamma(L)} \right]
\end{eqnarray}
and for $|h| \ge J$ as
\begin{eqnarray}\label{eq:ssfl}
\langle s \rangle^{sf}_l & = & \mathrm{sgn}(h) \frac{\cosh \widetilde{\beta} J - \mathrm{e}^{ \widetilde{\beta} |h|}}{\sinh  \widetilde{\beta} J - \mathrm{e}^{ \widetilde{\beta} |h|}} \left[ \frac{\Gamma\left( L, \frac{\ln N}{2} \right)}{\Gamma(L)} - \right. \nonumber \\
& & \left. \frac{\mathrm{e}^{-\widetilde{\beta} h} \sinh  \widetilde{\beta} J \Gamma\left(L,\mathrm{e}^{-\widetilde{\beta} h} \sinh  \widetilde{\beta} J \frac{\ln N}{2} \right)}{N^{\frac{-\mathrm{e}^{-\widetilde{\beta} h} \sinh  \widetilde{\beta} J-1}{2}}\Gamma(L)} \right]
\end{eqnarray}
A comparison of the theoretical predictions given by Eqs. (\ref{eq:ssfs}) and (\ref{eq:ssfl}) is shown in Fig. \ref{fig:ssf}a. It can be shown (Appendix \ref{app:b}) using analytical and numerical approach that the dependence of the crossover temperature $T_c$ on tree size $N$ is best described by
\begin{equation}\label{eq:sftc}
T_c \approx \frac{2}{1 + \frac{4}{3} \mathrm{W}\left( \frac{\ln N }{4 \mathrm{e}} \right)},
\end{equation}
where $\mathrm{W}(...)$ is Lambert W function. A comparison of the crossover temperature obtained by numerically solving Eq. (\ref{eq:ssf}) with the predictions of Eq. (\ref{eq:sftc}) is shown in Fig. \ref{fig:ssf}b. It is interesting to add here that for sufficiently large values of $x$ the function $\mathrm{W}(x)$ can be approximated with $\mathrm{W}(x) \approx \ln x - \ln \ln x$ which would suggest that for large values of $N$ the crossover temperature is given by $T_c \approx (2/3 \ln \ln N - \ln 2)^{-1}$.

\section{CONCLUSIONS}\label{sec:con}
In this paper we extended previously introduced model of the growing spin chain onto the case of deterministic and random scale-free trees. We have shown that for these topologies the analytical approach using Markov chain concept is still valid owing to the possibility of calculating the weighted spin on each level of the tree. Similarly to the chain case, the model exhibits a crossover temperature corresponding to maximal magnetization. Unlike the chain case, the crossover temperature decays very slowly [approximately as $(\ln \ln N)^{-1}$ compared to $(\ln N)^{-1}$ for the chain], which is connected to the fact of the effective diameter of the considered systems.

\begin{acknowledgments}
This work has been supported from Polish Ministry of Science and Higher Education (grant no. 0490/IP3/2011/71) as well as by the European Union in the framework of European Social Fund through the Warsaw University of Technology Development Programme, realized by the Center of Advanced Studies.
\end{acknowledgments}

\appendix
\section{DERIVATION OF THE CROSSOVER TEMPERATURE FOR DETERMINISTIC TREES}\label{app:a}
In order to get an analytical approximation of $T_c$ we use Eq. (\ref{eq:sts}) and assume that $\widetilde{\beta}h \ll 1$ which gives us the opportunity to set $\cosh \widetilde{\beta}h \approx 1$ and $\tanh \widetilde{\beta}h \approx \widetilde{\beta} h$:
\begin{equation}\label{eq:a2}
\langle s \rangle \approx  \widetilde{\beta} h \left\{ 1 + \frac{1}{N} - \frac{1 - \left[\left( 1- \mathrm{e}^{- \widetilde{\beta} J}  \right) z \right]^{L+1}}{N \left[1-\left(1 - \mathrm{e}^{- \widetilde{\beta} J}  \right) z \right]} \right\}
\end{equation}
Secondly, let us note that for $z^L \gg 1$ one can approximate Eq. (\ref{eq:nz}) with $N \approx z^L/(z-1)$. Making use of this fact and assuming $N \gg 1$ we get
\begin{equation}
\langle s \rangle \approx  \widetilde{\beta} h \left[ 1 +  \frac{(z-1) \left( 1- \mathrm{e}^{- \widetilde{\beta} J}  \right)^{L+1}}{1-\left(1 - \mathrm{e}^{- \widetilde{\beta} J}  \right) z} \right]
\end{equation}
Finally, also assuming that $\widetilde{\beta} J \gg 1$ we arrive at
\begin{equation}
\langle s \rangle \approx  \widetilde{\beta} h \left[ 1 +   \left( 1- \mathrm{e}^{- \widetilde{\beta} J}  \right)^{L+1} \right]
\end{equation}
It is interesting to observe here that this result does not depend on the branching factor $z$ what confirms the behavior seen in Fig. \ref{fig:tcLN}b.\\
As the next step we need to solve $\frac{\partial \langle s \rangle}{\partial T}=0$, i.e., 
\begin{equation}
\widetilde{\beta_c}(L+1) \mathrm{e}^{- \widetilde{\beta_c} J} \left(1 - \mathrm{e}^{- \widetilde{\beta_c} J} \right)^{L+1}=1+\left(1 - \mathrm{e}^{- \widetilde{\beta_c} J} \right)^{L}
\end{equation}
By taking the logarithm of both sides and assuming $L \gg 1$ as well as $\left(1 - \mathrm{e}^{- \widetilde{\beta_c} J} \right)^{L} \ll 1$ we arrive at
\begin{equation}
\widetilde{\beta_c} + L \mathrm{e}^{- \widetilde{\beta_c} J} = \ln \widetilde{\beta_c} L
\end{equation}
However, the above equation still fails to be solved by analytical methods. To overcome this problem, we use the approximation $\ln \widetilde{\beta_c} L = \ln (2L/k_B) - \ln T_c \approx \ln (\mathrm{e}L/k_B)$. Then, the resulting equation
\begin{equation}
\widetilde{\beta_c} + L \mathrm{e}^{- \widetilde{\beta_c} J} = \ln (\mathrm{e}L/k_B)
\end{equation}
has a solution of the form
\begin{equation}
T_c \approx \frac{2 J}{k_B \left[ \ln \frac{\mathrm{e} L}{k_B} + \mathrm{W}\left( - \frac{k_B}{\mathrm{e}}\right) \right]}
\end{equation}
Setting $J = k_B = 1$ leads us to the final result
\begin{equation}\label{eq:tcl}
T_c \approx \frac{2}{\ln L}.
\end{equation}

\section{DERIVATION OF THE CROSSOVER TEMPERATURE FOR RANDOM SCALE-FREE TREES}\label{app:b}
First, let us note that for $x \gg y$ we can write $\Gamma(x,y) \approx \Gamma(x)$ which gives us the opportunity to write Eq. (\ref{eq:ssfs}) as  
\begin{equation}\label{eq:b1}
\langle s \rangle \approx  \tanh  \widetilde{\beta} h \left[ 1 - (1-\mathrm{e}^{-\widetilde{\beta} J} \cosh  \widetilde{\beta} h) N^{-\frac{1}{2}\mathrm{e}^{-\widetilde{\beta} J} \cosh  \widetilde{\beta} h} \right]
\end{equation}
Secondly, as in the case of deterministic trees, we assume that $\widetilde{\beta}h \ll 1$ which results in setting $\cosh \widetilde{\beta}h \approx 1$ and $\tanh \widetilde{\beta}h \approx \widetilde{\beta} h$:
\begin{equation}\label{eq:b2}
\langle s \rangle \approx  \widetilde{\beta} h \left[ 1 - \left(1-\mathrm{e}^{-\widetilde{\beta} J} \right) N^{-\frac{1}{2}\mathrm{e}^{-\widetilde{\beta} J} } \right].
\end{equation}
Finally, assuming that $1 - \mathrm{e}^{-\widetilde{\beta} J} \approx 1$ and solving $\frac{\partial \langle s \rangle}{\partial T}=0$ we arrive at
\begin{equation}\label{eq:b3}
1-N^{-\frac{1}{2}\mathrm{e}^{-\widetilde{\beta_c} J}} \approx \frac{1}{2}\widetilde{\beta_c} J \mathrm{e}^{-\widetilde{\beta_c} J} N^{-\frac{1}{2}\mathrm{e}^{-\widetilde{\beta_c} J}} \ln N
\end{equation}
\begin{figure}
\centering
\includegraphics[width=\columnwidth]{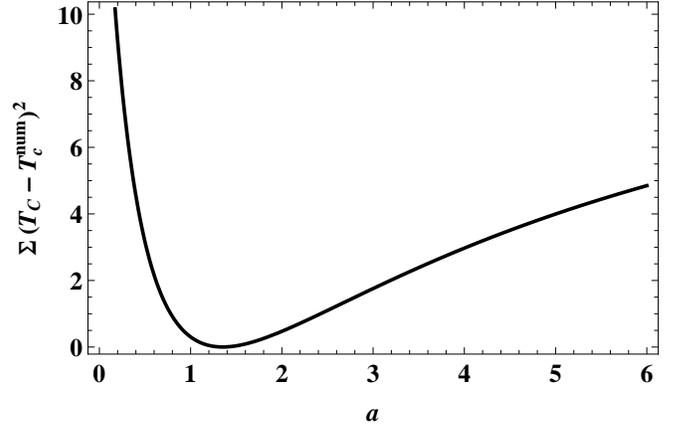}
\caption{Sum of the squares of displacement between the numerical and theoretical results of the crossover temperature for random scale-free trees versus parameter $a$ [see Eq. (\ref{eq:b7})].}
\label{fig:appb}
\end{figure}
At this point we use the fact that for $x \ll 1$ we can expand $N^{x/2}$ with Taylor series as $N^{x/2} \approx 1 + \frac{1}{2} \ln N x + \frac{1}{2} (\ln N x)^2$, which gives us the final equation
\begin{equation}\label{eq:b4}
\widetilde{\beta_c} J - \frac{1}{4} \ln N \mathrm{e}^{-\widetilde{\beta_c} J} \approx 1
\end{equation}
that has the solution
\begin{equation}\label{eq:b5}
T_c = \frac{2 J}{k_B \left[1 + \mathrm{W}\left( \frac{\ln N }{4 \mathrm{e}} \right) \right]}
\end{equation}
where $\mathrm{W}(...)$ is Lambert W function. Predictions of Eq. (\ref{eq:b5}) for $k_B = J = 1$ are shown in Fig. \ref{fig:ssf} with dashed line, suggesting divergence with the numerical solution $T^{num}_c$ of Eq. (\ref{eq:ssfs}), which is caused by the Taylor series expansion. To overcome this issue we propose the solution in a form
\begin{equation}\label{eq:b6}
T_c(a) = \frac{2}{1 + a \mathrm{W}\left( \frac{\ln N }{4 \mathrm{e}} \right)},
\end{equation}
where $a$ is chosen so that the sum
\begin{equation}\label{eq:b7}
\sum \left( T_c(a) - T^{num}_c \right)^2
\end{equation} 
is minimal. Numerical minimization of the above functional gives $a \approx 4/3$ (see Fig. \ref{fig:appb}).

\end{document}